\documentclass[3p,times]{elsarticle}

\usepackage{amssymb}
\usepackage[T1]{fontenc}       % Use modern font encodings
\usepackage{gensymb}
\usepackage{newunicodechar}
\usepackage{textcomp}
\usepackage{setspace}
\usepackage{graphicx}
\usepackage{float}
\usepackage{amsmath,environ}
\usepackage{stfloats}
\usepackage{lipsum}
\usepackage[figuresright]{rotating}

\begin{document}
	
	\begin{frontmatter}



\title{Grain Boundary Driven Plateau-Rayleigh Instability in Multilayer Nanocrystalline Thin Film: A Phase-field Study}


\author[1]{Tamoghna Chakrabarti\corref{cor1}}
\ead{tchakrabarti@mines.edu}
\author[2,4]{Nisha Verma}
\author[3,4]{Sukriti Manna\corref{cor1}}
\ead{smanna@mines.edu}

\address[1]{Department of Metallurgical and Materials Engineering, Colorado School of Mines, Golden, CO 80401, USA}
\address[2]{Department of Materials Science and Engineering, University of Illinois at Urbana-Champaign,  IL 61801, USA}
\address[3]{Department of Materials Engineering, Indian Institute of Science, Bangalore 560012, India}
\address[4]{Department of Mechanical Engineering, Colorado School of Mines, Golden, CO 80401, USA}
\cortext[cor1]{Corresponding authors}

\begin{abstract}
	Thermal stability of nanocrystalline multilayer thin film is of paramount importance as the applications often involve high temperature. Here we report on the layer instability phenomenon in binary polycrystalline thin film initiating from the grain boundary migrations at higher temperatures using phase-field simulations. Effect of layer thickness, bilayer spacing and the absence of grain boundary are also investigated along with the grain boundary mobility of individual phases on the layer stability. Layer instability in the polycrystalline film is shown to arise from the grain boundary grooving which originates spontaneously from the presence of grain boundaries. Our results show that the growth of the perturbation generated from the differential curvature follows Plateau-Rayleigh instability criterion. Increase in layer thickness, lower bilayer thickness as well as lower grain boundary mobility improve layer stability. Phase-field simulations show similar microstructural evolution as has been observed in our Zirconium (Zr)/Zirconium Nitride (ZrN) system experimentally. Detail analysis performed in this work to understand the mechanisms of layer instability leads us to predict measures which will improve the thermal stability of multilayer nanocrystalline thin film.
\end{abstract}


\end{frontmatter}

\section*{Highlights}
\begin{itemize}
	\item Instability in nanocrystalline multilayer thin film arises from grain boundary grooving followed by Plateau-Rayleigh instability
	
	\item Over a critical thickness which is dictated by Plateau-Rayleigh instability criteria layers remain stable
	
	\item Lower grain boundary mobility makes layer more stable
	
	\item Lower bi-layer thickness helps in the multilayer stability
\end{itemize}

\section{Introduction}
Multilayer thin films are characterized as an assemblage of two phases alternately stacked for desirable properties. They have been successfully applied as surface protection for tribological applications, diffusion barriers in microelectronics, hard wear resistant coatings on cutting tools due to their high hardness, fracture toughness, chemical inertness and thermal stability. \cite{miyata2015high,hultman2000thermal,xu2002microstructure,yu2015multilayer}. Essentially, the designing of an appropriate microstructure is the key for desired thin film properties. Multilayer shows better performance compared to single layer film. The design with metal/nitride multilayer coating show improved fracture resistance without much compromise of hardness \cite{shih1992ti,verma2013detailed,madan1996growth,madan1998enhanced,bozyazi2004comparison}, implying the desirable properties are an outcome of reduction in grain size and increased number of interfaces. 

The thermal stability is defined as deterioration of properties as a function of temperature, involving changes in microstructure,  grain growth, composition, phase transformation and oxidation. Some multilayer thin film coatings are used for tribological application owing to their superior mechanical properties. During tribological applications, temperature in the coated work piece can rise to very high temperature of around 1000\textcelsius. High temperature generated during the operation may lead to grain coarsening and disintegration of the layer structure, hence degrading the mechanical properties. Therefore, if multilayer were to replace monolayers due to superior mechanical properties, it is imperative to understand their stability at higher temperature, mainly by loss of nanostructure due to the migration of the grain boundaries. 

Pinching of the multilayer due to heat treatment has been observed in many different systems. In case of Silver (Ag)/ Silicon (Si) multilayer system with layer thickness between 3-5 nm, discontinuity in the layer has been observed after heat treatment above 600K \cite{kapta2003degradation}. They have also observed granular Ag at 723 K temperature. Thermal stability of Copper (Cu)-Tungsten (W) multilayer system with 5nm thickness has been studied by Moszner et.al. They have observed two main microstructural changes i.e. formation of line-shaped protrusions at the temperature higher than 500\textcelsius and degradation of the nanolayered structure at temperatures higher than 700\textcelsius. The initial multilayer structures completely disintegrate into a nanocomposite structure, comprising globular W particles embedded in a Cu matrix\cite{moszner2016thermal}.

\begin{figure}[H]
	\centering
	\includegraphics[width=3in]{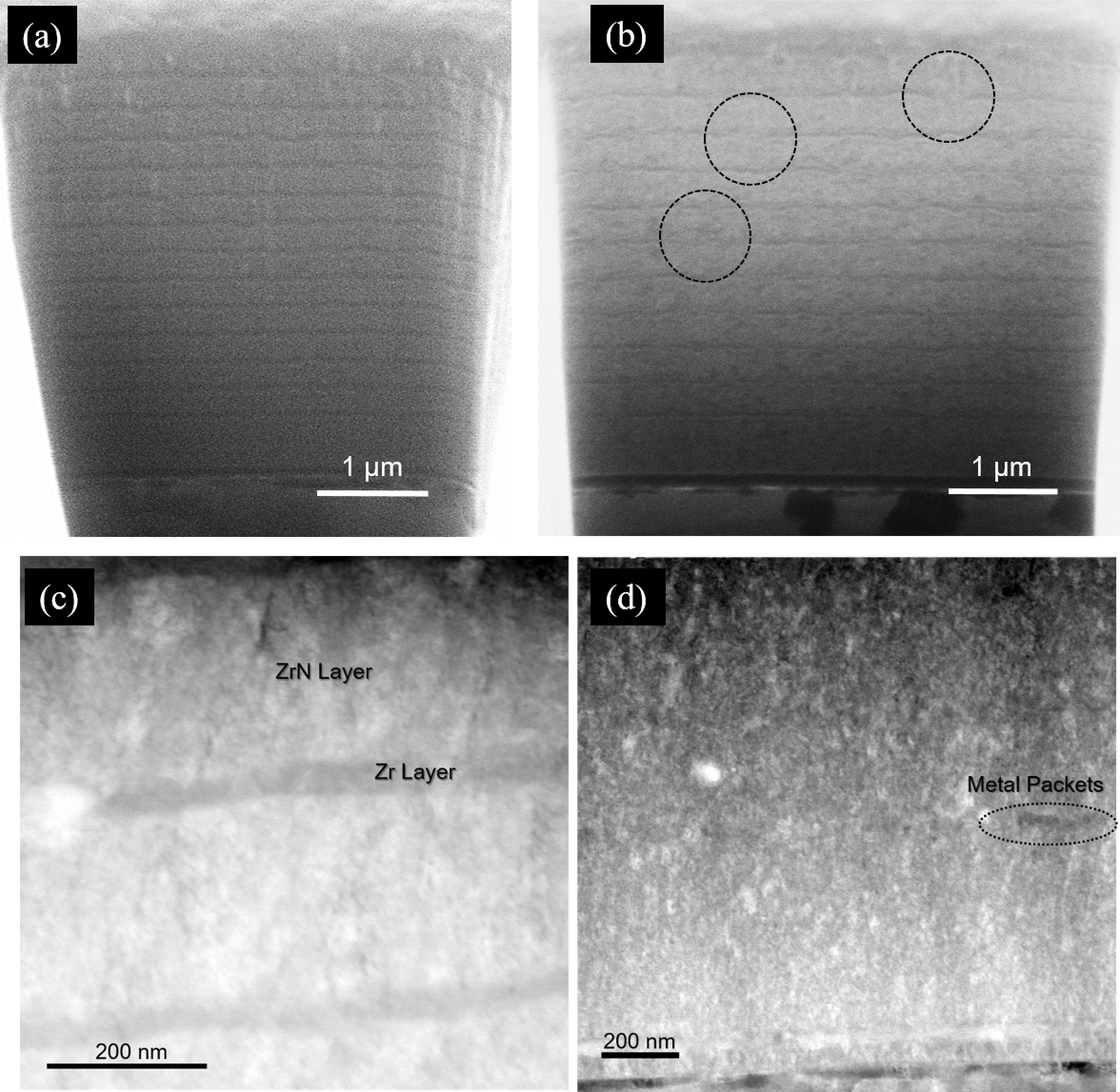}
	\caption{FIB cross sections showing layer structure at (a) 773K and (b) 973K for 250/50 nm multilayer coating. The interrupted layered structure is marked in (b). HAADF micrographs for as grown (c) and after 1173K (d), where the residual metal packets are marked in (d).
		}
		\label{fig_experiment}
\end{figure}

Our experimental Fig. \ref{fig_experiment} shows the microstructural evolution in Zirconium (Zr)/Zirconium Nitride (ZrN) multilayer thin film before and after the thermal treatment. The darker phases in Fig. \ref{fig_experiment} are Zr and brighter one are ZrN. The  Focused Ion Beam (FIB) cross section of the multilayer is shown in Fig. \ref{fig_experiment}a. Fig. \ref{fig_experiment}c shows the cross sectional high angle annular dark field (HAADF)-TEM micrographs  of the same multilayer before the heat treatment. Fig. \ref{fig_experiment}b represents the layer structure after the heat treatment 973K where it is evident that the continuity of the layer structure gets disrupted and the nitride grows through the metal layer not as a parallel front rather through some specific grains. Fig. \ref{fig_experiment}d shows that, even higher temperature heat treatment at 1173K completely disintegrates the layer structure. Residue of metal can be seen as few metal packets retained inside the nitride matrix after heat treatment at 1173K.
 
The above mentioned instability problem of nanocrystalline multilayer is in mesoscale length scale owing to the nanometer sized grains and layers. The mechanisms involved are atomic diffusion and grain boundary migration. Therefore, phase-field modeling technique is ideally suited to model this phenomenon \cite{mukherjee2011thermal}. It can also allow different grain boundary mobility in different phases, which can play an important part in the temporal evolution of the system. Additionally, this modeling technique correctly captures the effect of curvature which plays a significant role in producing the Plateau-Rayleigh instability leading to disintegration of the layer structure. Phase-field modelling of Plateau-Rayleigh instability in nano-wire and systems with continuous porous channel  has been studied previously \cite{joshi2016phase,wang2016detachment}. But the role of grain boundary migrations on the disintegration of nanocrystalline multilayer thin film has never been studied extensively through phase-field or any other computational modelling technique. 

The main goal of this work is to understand the underlying mechanism (or mechanisms) behind the layer instability that can arise in nanocrystalline multilayer thin film. Additionally, the effect of layer thickness, bilayer spacing and grain boundary mobility on the stability of the multilayer thin film are also extensively explored in this study. 

\section{Methods}

\subsection{Experimental}
Multilayer coating with bilayer spacing of 250/50 nm (ZrN/Zr) repeated for the total thickness of 5$\mu$m were magnetron sputtered at TEER coating, UK. Heat treatments were done at 773-1173K for 1hr, for the thermal stability study. Heat treatments were done in a vacuum furnace at the heating rate of 10$\degree$C/min with samples vacuum sealed in a quartz tube to further reduce the oxygen content. Microstructural characterization of the as-grown and heat treated samples were done by Focused Ion Beam (FIB, strata, FEI Inc., USA) and Transmission electron microscopy (TEM, Tecnai F-30, FEI Inc., USA). Cross-section imaging was used to image layers in secondary electron imaging mode in FIB. High angle annular dark field (HAADF) imaging was used to generate Z contrast images in TEM.

\subsection{Computational}
Our binary ($\alpha$ and $\beta$) polycrystalline multilayer system is described by three types of phase-field variables : $c$ describes the chemical composition,
$\eta_{i}^{\alpha}$ for $i = 1, 2,..., n_{\alpha}$ describes $n_{\alpha}$ unique grain orientations in the $\alpha$ phase and $\eta_{j}^{\beta}$ for $j = 1, 2,..., n_{j}^{\beta}$ describes $n_{\beta}$ unique
grain orientations in the $\beta$ phase. Chemical compositions of $\alpha$ and $\beta$ phases  are $c_{\alpha}$ and $c_{\beta}$ respectively. The temporal evolution of microstructure is governed by the Cahn-Hilliard \cite{cahn1961spinodal} and the Allen-Cahn equations \cite{allen1979microscopic}
\begin{equation}
	\frac{\partial c}{\partial t} = \nabla.M\nabla\mu
\end{equation}
\begin{equation}
	\frac{\partial \eta_{i}^{\alpha}}{\partial t} = -L_{\alpha}\frac{\delta (F/N_v)}{\delta \eta_{i}^{\alpha} }
\end{equation}
\begin{equation}
	\frac{\partial \eta_{j}^{\beta}}{\partial t} = -L_{\beta}\frac{\delta (F/N_v)}{\delta \eta_{j}^{\beta} }
\end{equation}
where $t$ is time, M is the atomic mobility (related to atomic diffusivity) 
\begin{equation}
	D =M\frac{\partial^2F}{\partial c^{2}} 
\end{equation}
where $F$ is the bulk free energy density defined below, $L$ ($L_\alpha$ and $L_\beta$) is a relaxation coefficient for order parameter field $\eta$ (grain boundary mobility) and $\mu$ is the chemical potential, defined as the variational derivative of the total free energy per lattice site, $F$, with respect to the local material density,
\begin{equation}
	\mu = \frac{\delta(F/N_v)}{\delta c}
\end{equation}
$F$ is described by,
\begin{equation}
\begin{split}
	F = N_{v}\int_{\Omega}[f(c,\eta_{i}^{\alpha},\eta_{i}^{\beta})+\kappa_c(\nabla c)^2 +  \sum_{i=1}^{n_\alpha} \kappa_{\eta}^\alpha(\nabla \eta_{i}^\alpha)^2 +  \sum_{i=1}^{n_\beta} \kappa_{\eta}^\beta(\nabla \eta_{i}^\beta)^2]d\Omega
	\end{split}
\end{equation}
Our free energy function is constructed from free energy functional described by Fan et.al. and Poulsen et. al. \cite{fan1998phase,poulsen2013three}. Free energy functional $f$ is described by,

\begin{equation}
	\begin{split}
		f(c,\eta_{i}^{\alpha},\eta_{i}^{\beta}) = \frac{A}{2} (c-c_m)^2 + \frac{B}{4}(c-c_m)^4 +  \frac{D_{\alpha}}{4}(c-c_{\alpha})^4 +\frac{D_{\beta}}{4}(c-c_{\beta})^4  
		-\frac{G_{\alpha}}{2}(c-c_{\beta})^2\sum_{i=1}^{n_{\alpha}}(\eta_{i}^{\alpha})^2 +  \\ \frac{d_{\alpha}}{4}\sum_{i=1}^{n_{\alpha}} ({\eta_{i}^{\alpha}})^4 
		-\frac{G_{\beta}}{2}(c-c_{\alpha})^2\sum_{i=1}^{n_{\beta}}(\eta_{i}^{\beta})^2 +  \frac{d_{\beta}}{4}\sum_{i=1}^{n_{\beta}} ({\eta_{i}^{\beta}})^4 
		+ \frac{\epsilon}{2} \sum_{i=1}^{n_{\alpha}}(\eta_{i}^{\alpha})^{2}\sum_{j=1}^{n_{\beta}}(\eta_{j}^{\beta})^{2} + \\ \epsilon_{\alpha}\sum_{i=1}^{n_{\alpha}}\sum_{j>i}^{n_{\alpha}}(\eta_{i}^{\alpha})^2(\eta_{j}^{\alpha})^2
		+ \epsilon_{\beta}\sum_{i=1}^{n_{\beta}}\sum_{j>i}^{n_{\beta}}(\eta_{i}^{\beta})^2(\eta_{j}^{\beta})^2
	\end{split}
\end{equation}

Here $c_m = (c_\alpha +c_\beta)/2$, $A, B, D_\alpha, D_\beta, G_\alpha, G_\beta, \epsilon, \epsilon_\alpha, \epsilon_\beta$ are constants. Dihedral angle between $\alpha$ phase and  $\beta$ phase grain boundaries in our model is 140$\degree$. We have used periodic boundary condition. Semi-implicit Fourier spectral method was used for solving kinetic equations \cite{zhu1999coarsening}. Discrete Fourier transforms were performed using FFTW software package \cite{frigo1998fftw}.

\begin{table}[H]
	\centering
	\caption{Parameters Used in Simulations}
	\label{tb0:comp_detail}
	\begin{tabular}{lc}
		\hline
		Parameters  & Non dimensional Values  \\
		\hline
		$C_\alpha$   & $0.05$   \\
		$C_\beta$ & $0.95$   \\
		$A$  & $2.00$   \\
		$B$  & $9.80$   \\		
		$D_\alpha, D_\beta$  & $1.52$   \\		
		$G_\alpha, G_\beta$  & $1.23$   \\		
		$d_\alpha, d_\beta$  & $1.00$ \\		
		$\epsilon, \epsilon_\alpha, \epsilon_\beta$  & $7.00$ \\		
		$\kappa_c$ & $0.75$\\
		$\kappa_\eta^\alpha$ & $1.25$\\
		$\kappa_\eta^\beta$ &  $1.00$\\ 
		$\Delta x, \Delta y$ & $2.00$\\
		$\Delta t $ & $0.20$\\		
		\hline
	\end{tabular}
\end{table}

\section{Results and Discussions}

\begin{figure}[H]
	\centering
	\includegraphics[width=3in]{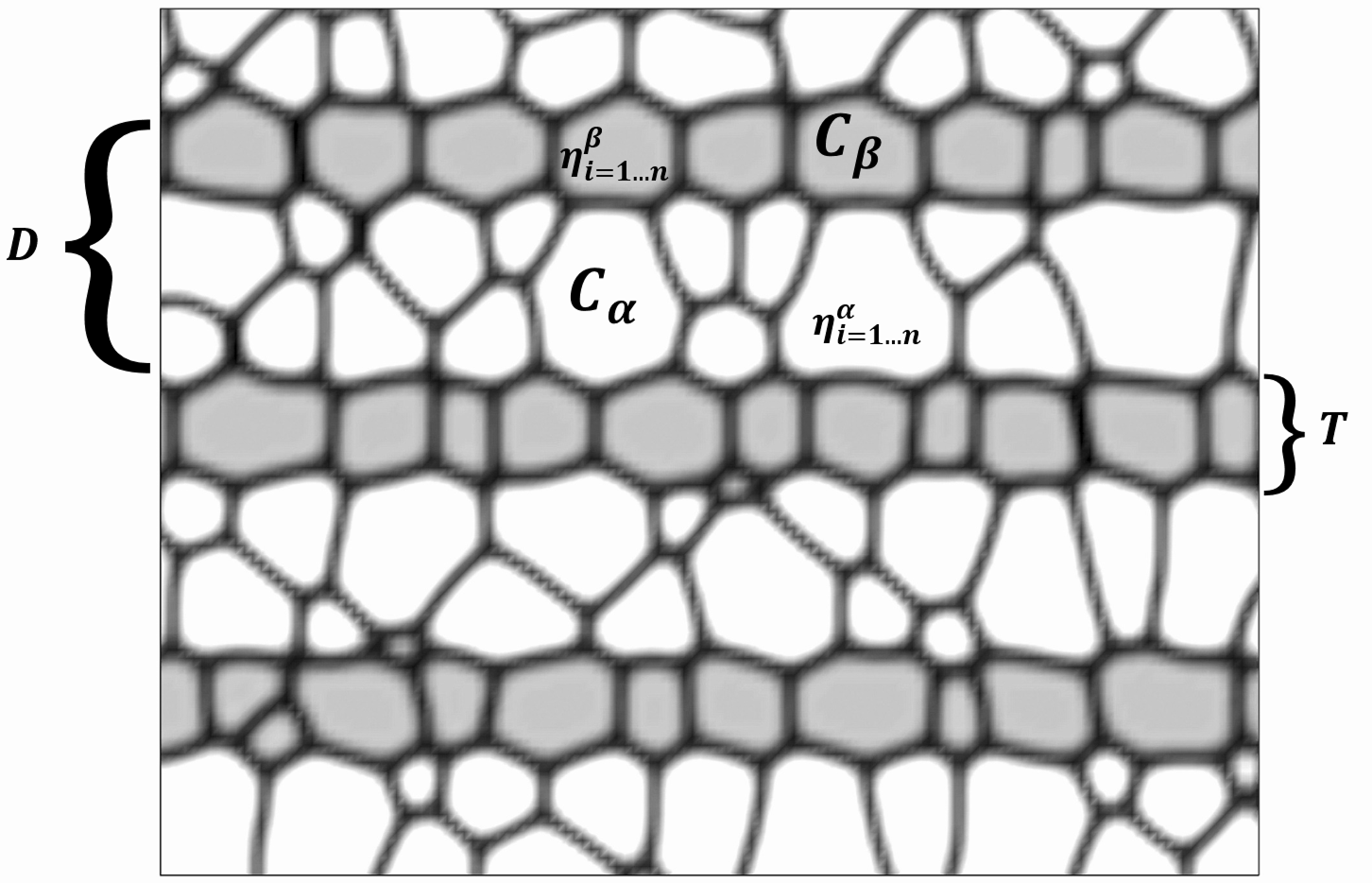}
	\caption{Two dimensional schematic of multi layer thin film, showing two different layer compositions and order parameters as well as $\beta$ layer thickness (T), and bilayer thickness (D).}
	\label{fig_Schematic}
\end{figure}

Fig. \ref{fig_Schematic} shows the schematic of the multilayer thin film studied in this work. $C_\alpha$ and $C_\alpha$ are the compositions of two alternate layers. We have not applied any external perturbation at $\alpha-\beta$ interface in this study. In general, our $\beta$ layer thickness is lower than that of the $\alpha$ layer.

\subsection{Mechanism of Film Disintegration}
\begin{figure}[H]
	\centering
	\includegraphics[width=4.0in]{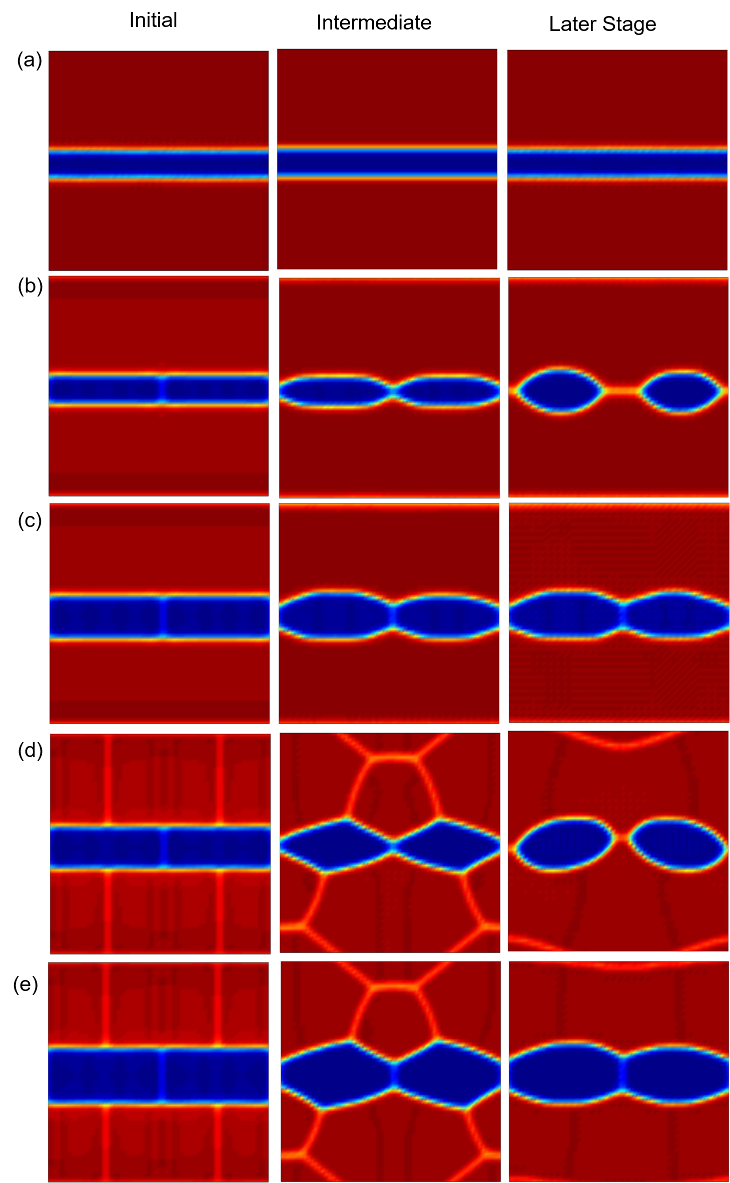}
	\caption{Early, intermediate and final stage of evolution in (a) single crystal multilayer thin film with T = 8, (b) polycrystalline layer embedded in single crystal layer of thickness T = 16 and (c) T = 24, (d) polycrystalline layer embedded in polycrystalline layer of thickness T = 24 and (e) T = 32. Here $\alpha$ and $\beta$ layers are described by red and blue colors respectively}. 
	\label{fig_small_size}
\end{figure}
To understand the underlying mechanism behind the film disintegration, we have studied an idealized case with 128x128 system size along with alternate layers of $\alpha$ and $\beta$ phase. We have varied the $\beta$ layer thickness (T) from 16 to 40.
\subsubsection{Case I: Both the layers are Amorphous (or single crystal)}
For all the thickness range of $\beta$ layer, no rupture in $\beta$ layer is observed when both the layers are amorphous or single crystal (Fig.  \ref{fig_small_size}a). This suggests that when the layers are devoid of grain boundaries they remain stable. We have not imposed any perturbation in this study.
\subsubsection{Case II: Presence of Grain Boundaries in one ( $\alpha$ or $\beta$) layer }
In this situation, we have observed rupture in $\beta$ layer at the later stage when T = 16 (Fig. \ref{fig_small_size}b). The presence of  grain boundary in $\beta$ layer causes grain grooving which leads to  perturbation in the  $\alpha$-$\beta$ interface. This perturbation grows over time and finally ruptures the $\beta$ layer. Thus the $\alpha$ phase grows  between two $\beta$ grains at the later stage. Therefore, the presence of grain boundary in $\beta$ layer (or $\alpha$) makes it less stable than Case I. But, Fig. \ref{fig_small_size}c, shows at a higher thickness of (T = 24), the $\beta$ 
layer remains stable.
\subsubsection{Case III: Presence of Grain Boundaries in both $\alpha$ and $\beta$ layer}
\begin{figure}[H]
	\centering
	\includegraphics[width=4in]{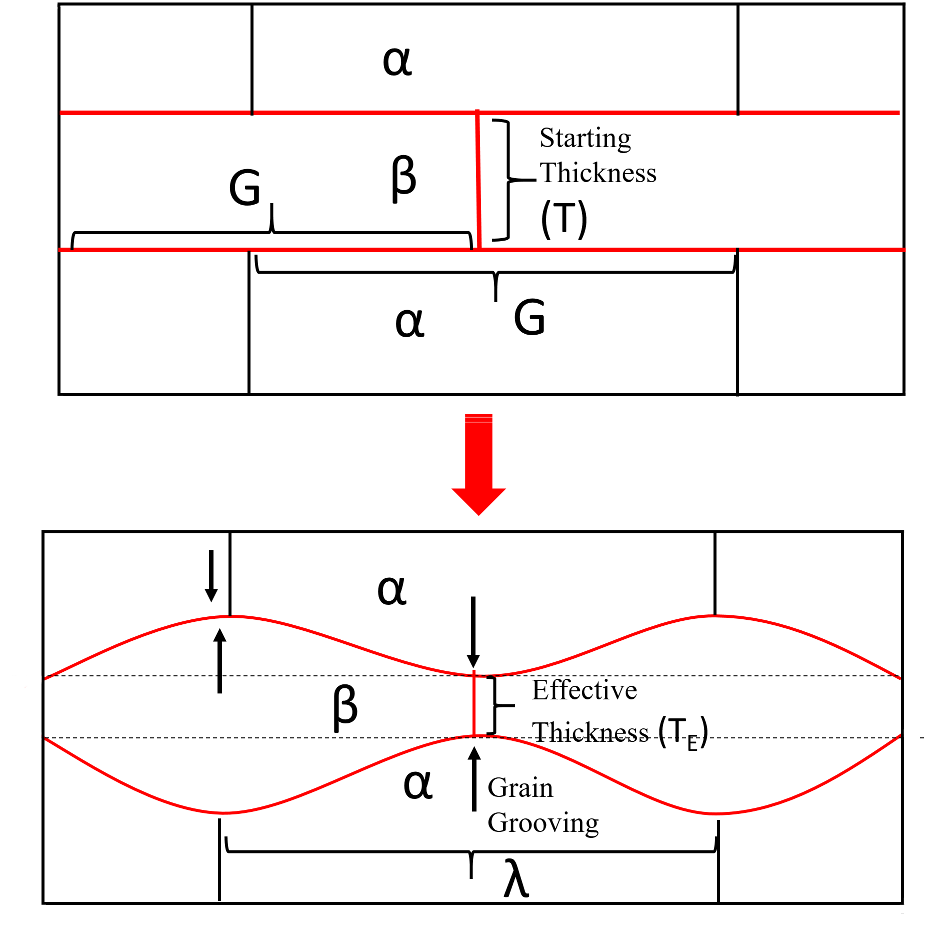}
	\caption{Schematics of early and final stages of multilayer thin film. Arrows showing the direction of grain grooving.}
	\label{fig_mech_schematic}
\end{figure}

Fig. \ref{fig_small_size}d, shows the result of a polycrystalline $\beta$ layer of (T = 24) embedded in polycrystalline $\alpha$. The grain boundary at the middle of the $\beta$ layer causes grain grooving just as before, but the grain boundary of $\alpha$  phase also experiences grain grooving in reverse directions. The presence of grain boundaries in both $\alpha$ and $\beta$ phases makes it more unstable (Fig. \ref{fig_small_size}d) than the previous two cases as rupture of $\beta$ layer is observed even at T = 24. This can be explained by Plateau-Rayleigh (PR) instability \cite{strutt1878instability}.
\begin{eqnarray}
KT_E <1\\
K=\frac{2\pi}{\lambda}\\
\lambda=G \label{eq_G}\\
T_E<\frac{\lambda}{2\pi}\\
T_E<\frac{G}{2\pi} \label{eq_PR}
\end{eqnarray}
According to PR instability, the layer becomes unstable when K$T_E$ < 1, where K is the wave number and $T_E$ is the effective thickness of the $\beta$ layer and G is grain size. Due to grain grooving, effective thickness $T_E$ is always lower than the starting thickness T. For our simulation parameters, PR instability will make the layer unstable when $T_E$ < 10 as shown by equation \ref{eq_PR}. Fig. \ref{fig_mech_schematic} shows the corresponding schematics which elaborates the formation of sinusoidal perturbation due to the grain grooving. Owing to grooving on the both side of the $\beta$ layer, $T_E$ is 8 (<10) in the case of T = 16, 24 and layer ruptures. We observe that the layer remains stable when $T_E$ > 10 in the case of T = 32. Therefore, our simulations corroborate with the Plateau-Rayleigh instability criterion. 

Polycrystallinity in the two phases also affects the kinetics of the disintegration of $\beta$ layer. For the polycrystalline film embedded in a single crystal of thickness (T = 16) (Fig. \ref{fig_small_size}b), takes much longer time to rupture (Time = 1800), compared to polycrystalline film embedded in polycrystalline layer (Time = 600) of the same thickness. Polycrystalline $\alpha$ and $\beta$ film of T = 16 takes  around three times lesser time (Time = 600) to rupture than film of thickness T = 24 (Time = 1800).

\subsection{Factors Affecting the Stability of Thin Film :2D System}
From here onwards, we construct our discussion the evolution of 2D multilayer thin film with the system size of 1024x1024.

\subsubsection{Microstructural Evolution with Time}
\begin{figure}[H]
	\centering
	\includegraphics[width=2.5in]{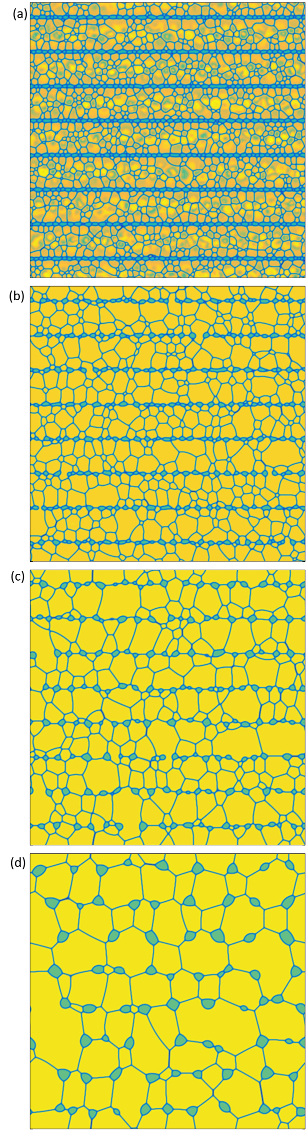}
	\caption{Microstructural evolution of multilayer thin film at (a) Time = 100, (b) Time = 500, (c) Time = 1700 and (d) Time = 10000.}
	\label{fig_time_ev}
\end{figure}

We will start the result section with the microstructural evolution of the multilayer thin films. Fig. \ref{fig_time_ev} a-d show the snapshots of the film disintegration for T = 16. The $\beta$ layer starts to pinch of at Time = 200. The $\alpha$ phase grows through the $\beta$ layer. This microstructure looks similar to the microstructure we have observed in Zr/ZrN multilayer after the heat treatment at 973K shown in Fig. \ref{fig_experiment}b. With the evolution of time, the $\beta$ layer gets visibly more disintegrated as shown in Fig. \ref{fig_time_ev}c. Finally, the layers become isolated islands of $\beta$ phase in Fig. \ref{fig_time_ev}d. Similarly, We observe similar isolated pockets of Zr metal after heat treatment at 1173K (Fig. \ref{fig_experiment}d). Higher $\beta$ layer thickness slows down the kinetics but the microstructural evolution trajectory remains same as seen for T = 16.
\subsubsection{Effect of Film Thickness}
\begin{figure}[H]
	\centering
	\includegraphics[width=6in]{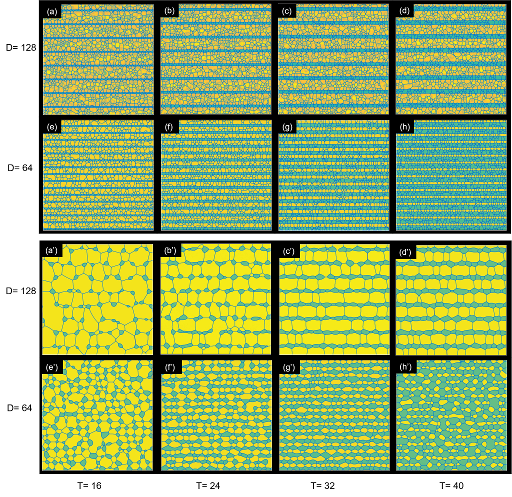}
	\caption{Early stage (Time = 100) and final stage (Time = 10000) microstructures of multilayer thin film with different layer thickness (T), and bi-layer thickness (D).}
		\label{fig_2D}
\end{figure}

In this section the effect of film thickness has been studied. Fig. \ref{fig_2D} a-g show the initial configuration whereas Fig. \ref{fig_2D} a'-g' show the final configuration after Time = 10000. In the Fig. \ref{fig_2D} a-d, from left to right $\beta$ layer thickness increases from 16 to 40. Final configurations a'-c' show that as we move from T = 16 to T = 32 the extent of $\beta$ layer disintegrity decreases. At T = 16, the $\beta$ film layer is least stable whereas when T = 40, $\beta$ layer is totally stable. For T = 16, The $\beta$ phase exists as droplet shape in grain boundaries or grain boundary triple points of $\alpha$ phase. Similar dependence of layer stability  on layer thickness has been shown experimentally by Misra et. al. in nanocrystalline Copper (Cu)-Niobium (Nb) multilayer system\cite{misra2005effects}.
As with the small system with idealized configuration described in Fig. \ref{fig_small_size}, here also the kinetics of film disintegration is controlled by layer thickness. Thicker film layer takes significantly longer time to break. Table \ref{tbl:film_thickness} shows the film thickness and the corresponding time to break.

\begin{table}
	\centering
	\caption{$\beta$ layer thickness vs Time to Disintegrate}
	\label{tbl:film_thickness}
	\begin{tabular}{lcc}
		\hline
		Film Thickness  & Time (D = 128) & Time (D = 64)  \\
		\hline
		T = 16   & 200  & 300   \\
		T = 24 & 1000  & 2200 \\
		T = 32  & 3500  & 8400 \\
		T = 40 & Does not break &  $\alpha$ Layer breaks\\
		\hline
	\end{tabular}
\end{table}

\begin{figure}[H]
	\centering
	\includegraphics[width=3in]{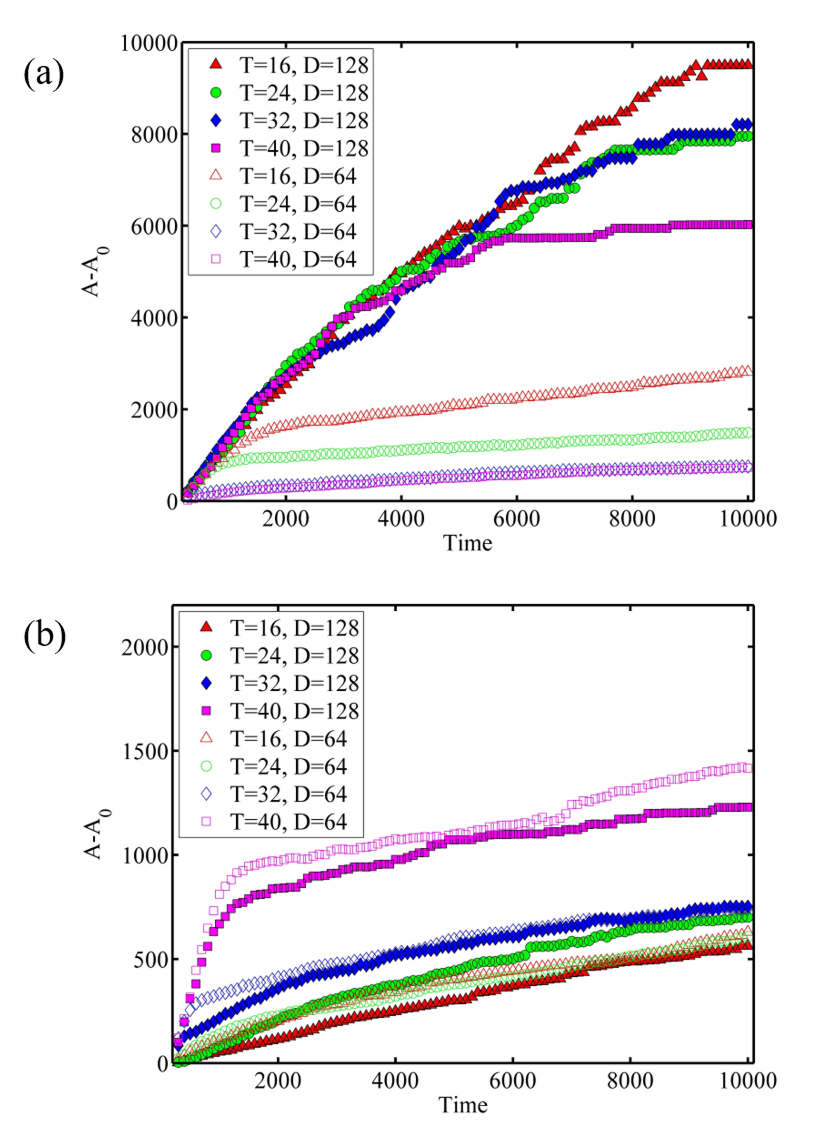}
	\caption{Average grain area in (a) $\alpha$ layer and (a) $\beta$ layer vs time plot for different $\beta$ layer thickness (T) and bi-layer thickness (D).}
	\label{fig_HK_layer}
\end{figure}

We have also studied the coarsening kinetics of the grains in individual $\alpha$ and $\beta$ phase. Fig. \ref{fig_HK_layer} a and \ref{fig_HK_layer} b show the grain coarsening with time in $\alpha$ and $\beta$ layer respectively. Grain size has been computed using Hoshen-Kopelman algorithm \cite{hoshen1976percolation} at each time step. As, $\alpha$ layer thickness decreases with increasing $\beta$ layer thickness (T), correspondingly we observe a monotonous decrease in final $\alpha$ grain area. $\alpha$ and $\beta$ grains only grow as large as the corresponding layer thickness, therefore, the grain size is dictated by their layer thickness. In all the four cases (T = 16, 24, 32 and 40), the grain area vs time plot of $\alpha$ phase shows two distinct stages. Initially, grains coarsen at much higher rate and then slows down significantly. This is because initially grains coarsen freely inside each of the individual layer without much hindrance as the number of the grain boundaries pinned by the grain grooving are relatively smaller than the total number of the grain boundaries. At this initial stage, the grain coarsening is independent of the layer thickness and coarsening rate remains the same in all the four cases. At later stage, the morphology of the grains changes to columnar. The grain boundaries of those columnar grains are pinned by the grain grooves on both the sides. Mechanisms of grain boundary pinning by the grooves are discussed by Mullins et.al.\cite{mullins1958effect} This pinning effect slows down the grain coarsening rate significantly. Therefore, we observe a crossover to much slower coarsening regime in the Fig. \ref{fig_HK_layer} a. As, the $\alpha$ layer is thinner with increasing $\beta$ layer thickness (T), the crossover point to columnar grains arrive sooner. Consequently, the crossover point between the faster and slower coarsening regime is shifted to the left.

Fig. \ref{fig_HK_layer} b shows the grain coarsening behavior with time in $\beta$ layer. As in $\alpha$ layer, $\beta$ layer grain area shows monotonous increase with increasing layer thickness. Here, in case of T = 16, 24 and 32, grain area vs time plot do not show two distinct regimes. Because of its lower thickness grain coarsening produces columnar grains in $\beta$ layer earlier than $\alpha$ layer (before time 200). Therefore, the fast coarsening regime can not be seen. But, for T = 40, as the thickness is higher,  a fast coarsening regime can be observed which is analogous to the  coarsening in the $\alpha$ layer.

\subsubsection{Formation of Columnar Grains}
\begin{figure}[H]
	\centering
	\includegraphics[width=2in]{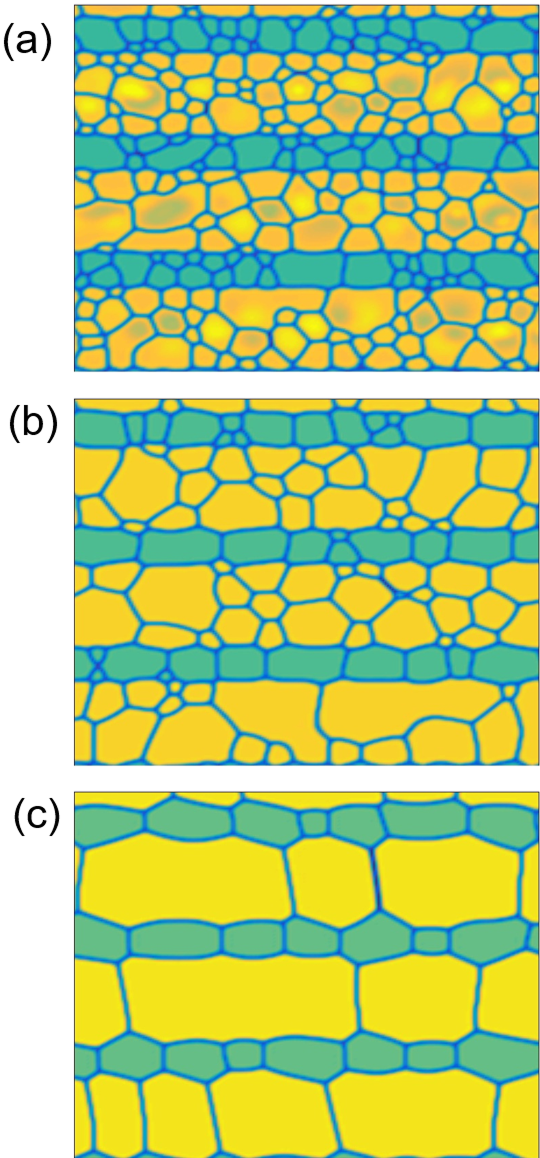}
	\caption{Micrograph showing evolution of grain structures with time, (a) Time = 100, (b) Time = 400 and  (c) Time = 4000 . Here yellow and green colored grains represents $\alpha$ and $\beta$ layers respectively. }
	\label{fig_columnar}
\end{figure}

Due to the slower movement of some grain boundaries (pinned at $\alpha-\beta$ interface by grain grooving) and faster grain boundary movement elsewhere, columnar grains will always arise in thin film eventually, provided the non-pinned boundaries can move. This observation can explain the reason behind the columnar grains formation in the nano-crystalline thin film produced experimentally \cite{thornton1974influence}. The sequence of the grain boundary migration and columnar grain formation is shown in Fig. \ref{fig_columnar}. Columnar grain arises in both  $\alpha$ and $\beta$ layer. It also can be observed that the columnar grain arises much earlier in  $\beta$ layer owing to lower film thickness.

\subsubsection{Effect of Bi-Layer Thickness (D)}
Here we have discussed the effect D on the stability of $\beta$ layer. In the Fig. \ref{fig_2D}, from left to right $\beta$ layer thickness increases from 16 to 40 and top row shows result from the bi-layer thickness of 128 and bottom layer shows the result from the bi-layer thickness of 64.  The initial configuration is shown in Fig. \ref{fig_2D} a-g whereas Fig. \ref{fig_2D} a'-g' show the final configuration after Time = 10000.

In the Fig. \ref{fig_2D}, top and bottom micrographs show the result of the same $\beta$ layer thickness but with different bi-layer thickness. Higher D value means higher  $\alpha$ layer thickness and less number of layers. For higher D values, the time to layer disintegration is lower for the same $\beta$ layer thickness T which is shown in Table \ref{tbl:film_thickness}. This can be explained through PR instability. For higher D,  $\alpha$ layer thickness is higher and resultantly, grain sizes are much larger as shown in Fig. \ref{fig_HK_layer} a. This corresponds to the higher wavelength ($\lambda$) of perturbation as shown by Eq. \ref{eq_G}. For the same $\beta$ layer thickness, higher $\lambda$ leads to the faster disintegration of the $\beta$ layer \cite{joshi2016phase}. Therefore, higher bi-layer thickness, by virtue of larger $\alpha$ phase grain size and correspondingly higher $\lambda$ leads to the faster disintegration of the multilayer. In the case of the T = 40 and D = 64, $\alpha$ layer becomes discontinuous as the $\alpha$  layer thickness (24) is lower than the $\beta$ layer.

Fig. \ref{fig_HK_layer} a, b show the grain coarsening with time in $\alpha$ and $\beta$ layer respectively. As, $\alpha$ layer thickness increases with higher bi-layer thickness (D), final grain area increases. In all the four cases (i.e. T = 16, 24, 32, and 40), the grain area vs time plot shows similar trends. Fig. \ref{fig_HK_layer} b shows the grain coarsening with time in $\beta$ layer. For $\beta$ layer, thicknesses (T) is the same for D = 64 and 128.  Here, grain area vs time shows similar trajectory in both the cases.

\subsubsection{Effect of Grain Boundary Mobility (L)}
In this section, we have discussed the effect of grain boundary mobility (L) on the stability of the multilayer thin film. Table \ref{tbl:film_L} shows the $L_\alpha$ and $L_\beta$ and the corresponding time to break. For the multilayer film with $\beta$ layer thickness T = 16 and 24 time to break increases with lower $L_\alpha$ and $L_\beta$. Instability in this film originates from the grain grooves. Formation of grain grooves depends on the movement of grain boundaries which is controlled by the parameter L. Therefore, it is natural that lower values of L ($L_\alpha$ and $L_\beta$) hinders the kinetics of film disintegration. 

Lowering of  $\alpha$ phase grain boundary mobility $L_\alpha$ keeping $L_\beta$ constant affects the time to breakage of $\beta$ layer more than the vice-versa for both T = 16 and 24. Higher  $\alpha$ phase grain boundary mobility $L_\alpha$, means larger grain size of $\alpha$ which increases the $\lambda$ (as shown by Eq. \ref{eq_G}) and leads to the faster disintegration of the $\beta$ layer. This effect is similar to the effect of larger D value discussed previously.  

\begin{table}[H]
	\centering
	\caption{Grain boundary mobility (L) vs Film Disintegration Time}
	\label{tbl:film_L}
	\begin{tabular}{lcc}
		\hline
		Film Thickness  & Time (T = 16) & Time (T = 24)  \\
		\hline
		L$_\alpha$ = 1.00, L$_\beta$ = 1.00   & 200  & 1000   \\
		L$_\alpha$ = 0.10, L$_\beta$ = 0.10 & 300  & 1600 \\
		L$_\alpha$ = 0.01, L$_\beta$ = 0.01  & 800  & 3800 \\
		L$_\alpha$ = 1.00, L$_\beta$ = 0.01 & 500 &  2500\\
		L$_\alpha$ = 0.01, L$_\beta$ = 1.00 & 700 &  3800\\
		\hline
	\end{tabular}
\end{table}

\begin{figure}[H]
	\centering
	\includegraphics[width=4in]{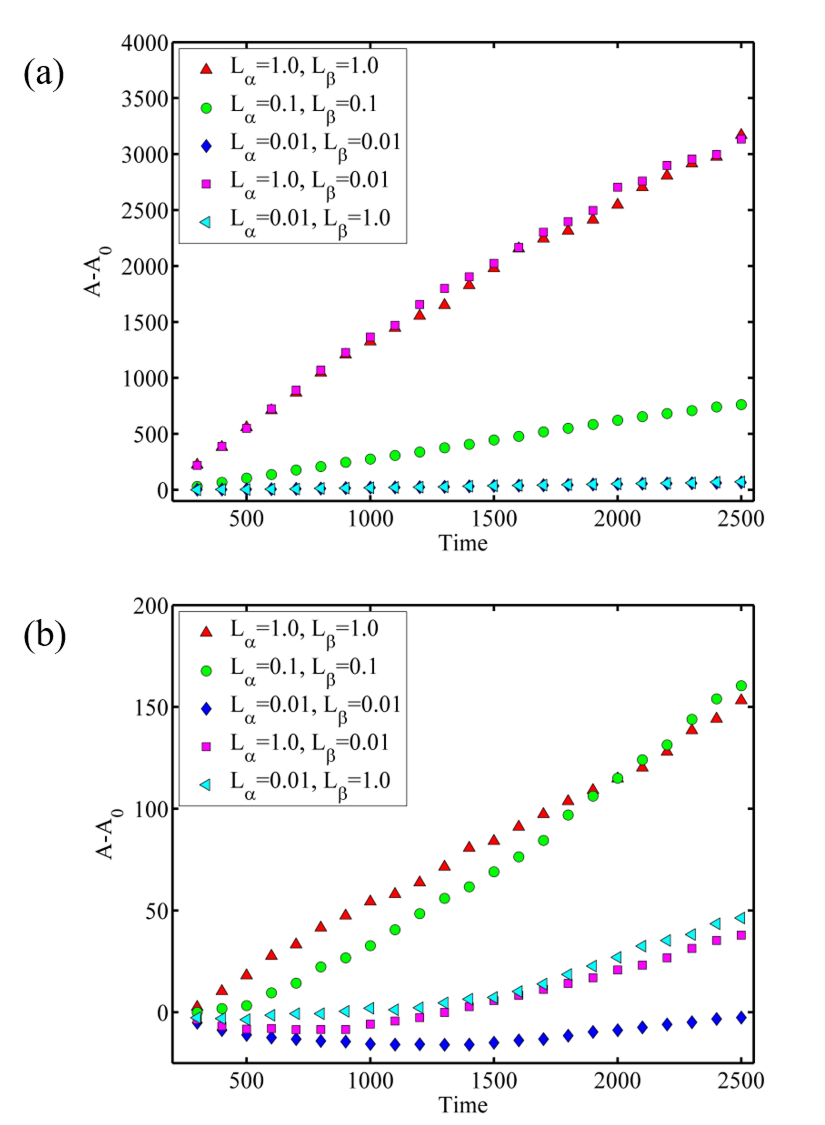}
	\caption{Average grain area in (a) $\alpha$ and (b) $\beta$ layer vs time plot for T = 16 and different L.}
	\label{fig_HK_T8}
\end{figure}

\begin{figure}[H]
	\centering
	\includegraphics[width=4in]{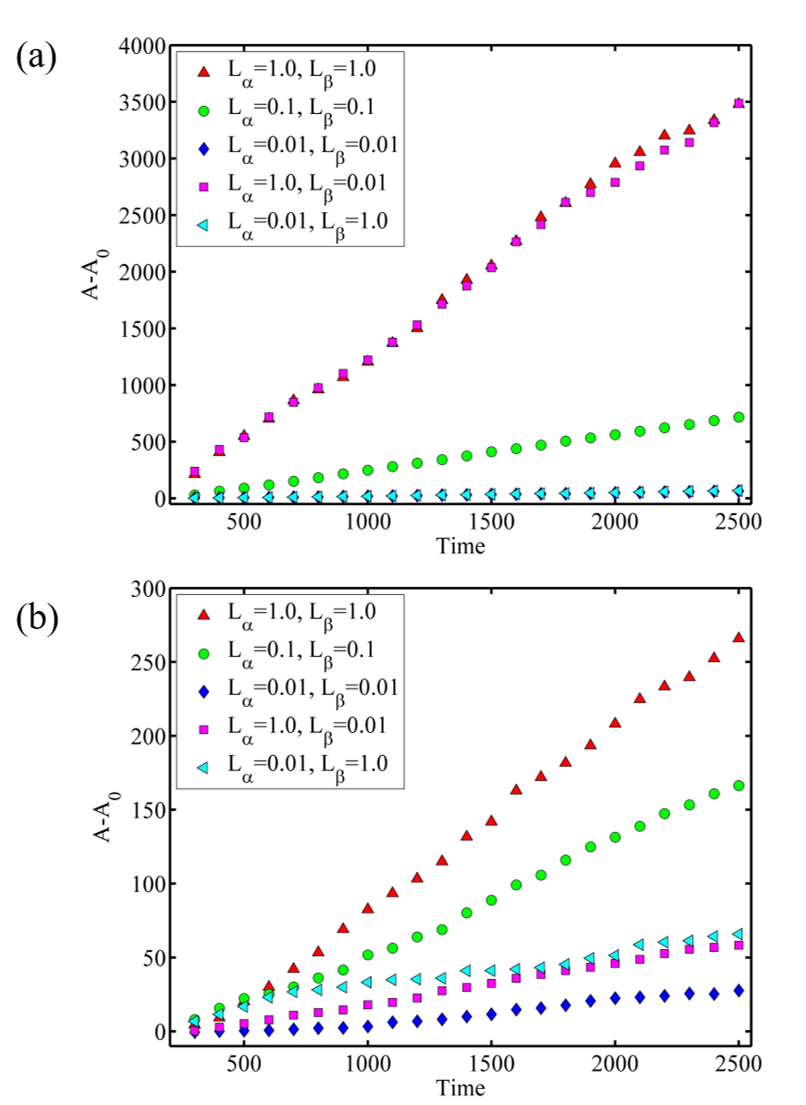}
	\caption{Average grain area in (a) $\alpha$ and (b) $\beta$layer vs time plot for T = 24 and different L.}
	\label{fig_HK_T12}
\end{figure}

Fig. \ref{fig_HK_T8} and \ref{fig_HK_T12} show the grain coarsening with time in $\alpha$ and $\beta$ layer and T = 16 and T = 24 respectively. $\alpha$ and $\beta$ layer grain size increases with higher values of respective L. In all the cases (i.e. T = 16 and 24), the grain area vs time plot shows a similar trend. $\alpha$ layer grain size is more dependent on L$_\alpha$ compared to the $\beta$ phase on L$_\beta$. This is because the grain boundaries in the $\beta$ phase are constrained by grain grooving from very early stage owing to lower thickness and therefore controlled by movement of pinned boundary which depends on bulk diffusion in addition to L values and moves at a significantly slower rate. 

Multilayer thin films in use today, mostly contain alternate layers of metal and ceramic. Our simulation results show that the metal layer due to its higher  grain boundary mobility (L), will be more susceptible to disintegration than the ceramic layer. Our model also indicates that the grain boundary mobility (L) of the ceramic phase has strong influence on the rupture of the metal layer. Faster the L is in ceramic layer, faster the metal layer disintegrates. 

To avoid the disintegration of the metal layer, it should be as thick as possible. Also, the bi-layer thickness should be as low as possible without causing disruption of the ceramic layer. This is because lower bilayer thickness with the higher thickness  of metal means lower thickness of ceramic layer. The ceramic layer by virtue of its much lower L, can resist the disintegration even at a thickness much lower than the metal layer.

The ceramic layer also should have as low grain boundary mobility as possible as the grain boundary mobility of ceramic layer affects the metal layer disintegration. As grain boundary mobility is a strong function of melting point, higher melting point ceramic will be better for multilayer stability. Similarly, the constituent metal of the metal layer also should have a higher melting point (i.e. lower grain boundary mobility) to achieve higher thermal stability.

\subsubsection{Presence of Amorphous Layer}
\begin{figure}[H]
	\centering
	\includegraphics[width=4in]{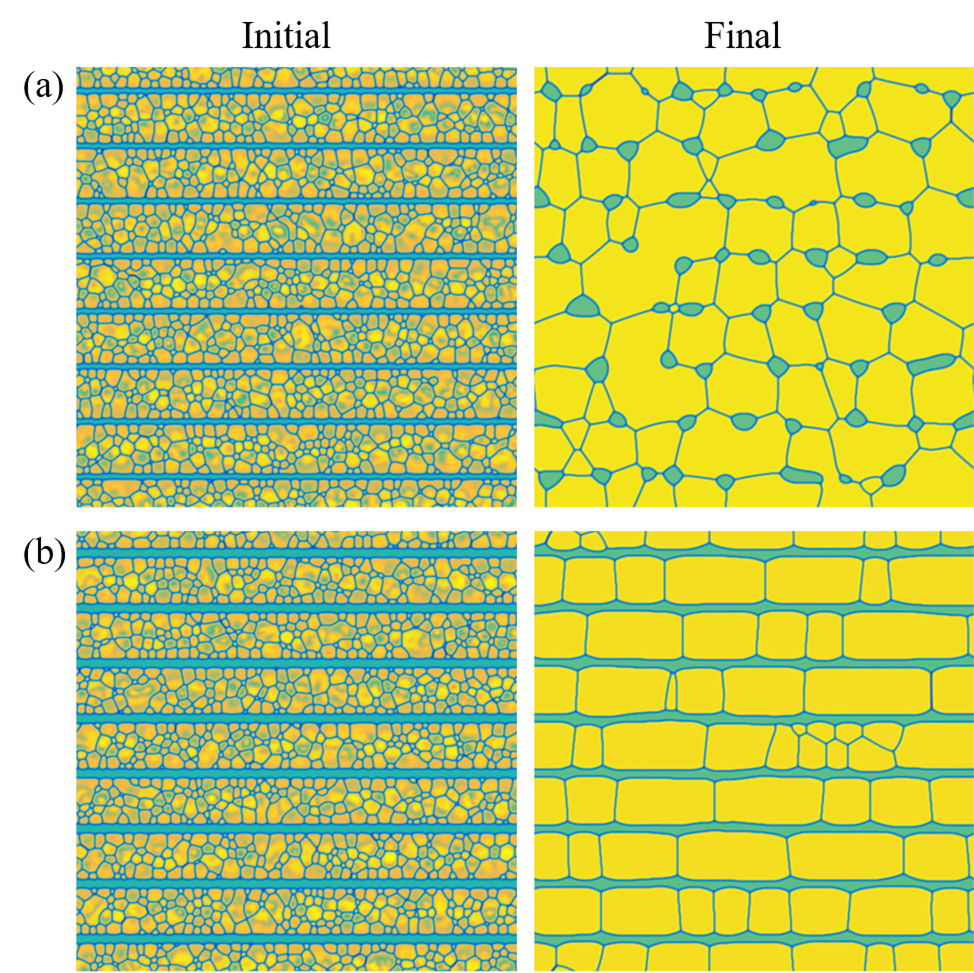}
	\caption{Initial (Time = 0) and later stage microstructures of multilayer thin film with different layer thickness is shown in (a) T = 16, and (b) T = 24}
	\label{fig_Amorphous}
\end{figure}
In multilayer coating, sometimes amorphous layer is formed due to kinetic reason \cite{singh2010residual}. Therefore, an important aspect of this study is their thermal stability. In our model, amorphous layers are those which do not have any grain boundaries. The same can be attributed to single crystal layer as well. Basically, in our model, there are no differences between amorphous and single crystal layer. The result of microstructural evolution in those systems has been shown in Fig. \ref{fig_Amorphous}. Here, we have discussed the microstructural evolution of the alternate amorphous and polycrystalline layer. If both the layers are amorphous, our small system size study has already shown that layers are stable for any thickness as without the grain boundaries and corresponding grooving there are no differential curvature which can lead to disintegration of the layer.
We have kept D to 128. The Fig. \ref{fig_Amorphous} shows that rupture of the multilayer occurs for T = 16 but not for T = 24 till the simulation time of 10000. We have observed the same in our small system size simulations as well. Time to rupture in this case i.e. amorphous $\beta$ layer is 600 which is three times longer than in the case of the polycrystalline $\beta$ layer. Clearly, grain boundaries in polycrystalline layer make rupture easier and faster. Slower disintegration of single crystal gold nanowire due to the PR instability has been observed experimentally as well\cite{karim2007influence}. Additional curvature from polycrystalline material due to grain grooving is likely responsible for such phenomenon as shown by our modelling results.

\subsection{Stability of Thin Film : 3D System}
\begin{figure}[H]
	\centering
	\includegraphics[width=4in]{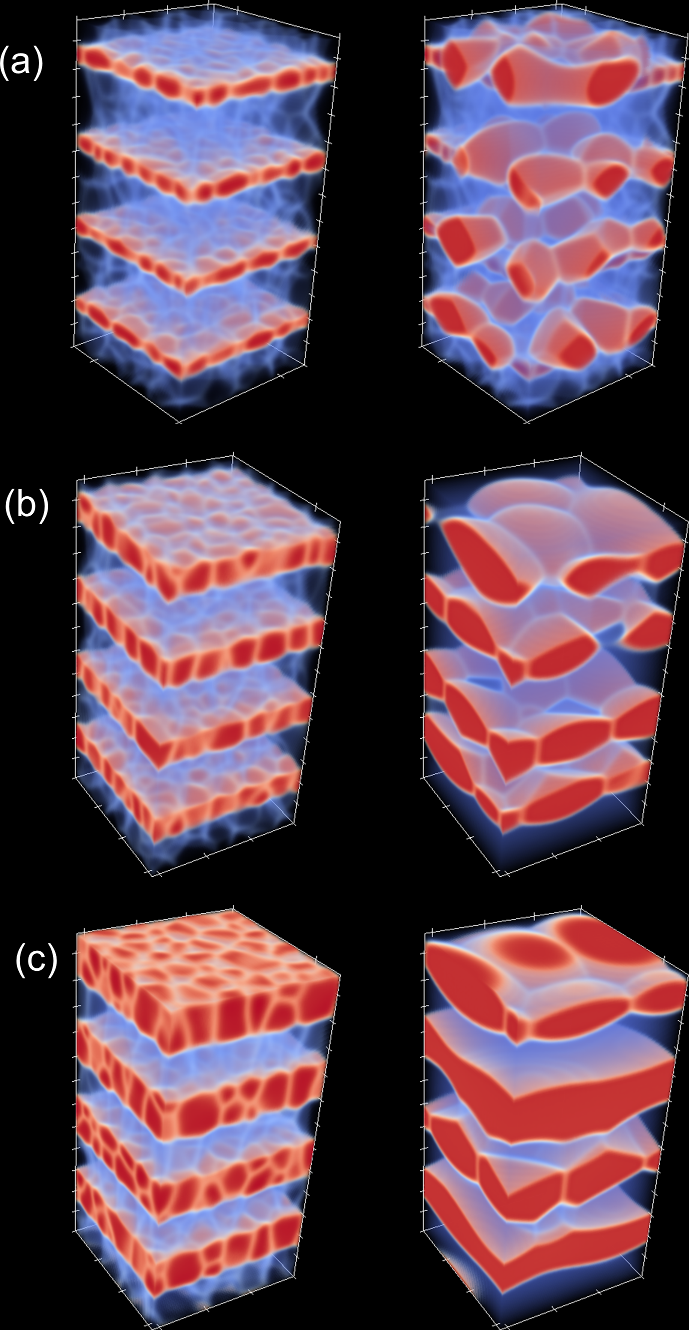}
	\caption{Initial (Time = 0) and later stage microstructures of multilayer thin film with different layer thickness is shown in (a) T = 16, (b) T = 24 and (c) T = 32.  $\beta$ layers are shown in red. In the  $\alpha$ layers only grain boundaries are shown for clarity.}
	\label{fig_3D}
\end{figure}
The result of the  phase field simulation of multilayer thin film in 3D is shown in Fig. \ref{fig_3D}. For 3D simulation system, the size of 256x128x128 is taken. Bilayer thickness (D) is fixed to 64 and the thickness of $\beta$ layer is varied from 16-32.  Fig. \ref{fig_3D} shows rupture in thin film in the case of T = 16 and 24. For T = 32, any rupture is not seen up to the time of 6000. In the case of T = 16, layer rupture starts at time 100. Time taken to rupture is longer in the case of T = 24 i.e. 1300. In comparison to the 2D system time taken to rupture in multilayer thin film is lower in the 3D system. Additional curvature in 3D system is likely responsible for such a reduction. Layer rupture is not observed for T = 32, till the simulation time of 5000. Due to the computational resource constraint, the longer simulation was not tried. Incidentally, for the same configuration, 2D simulation took 8400 time to show first sign of the rupture. Significant grain coarsening and columnar grain formation is observed in all the above-mentioned cases which is analogous to the result we have seen in the 2D system.

\section{Conclusion}
Our results provide a detail insight behind the instability phenomenon observed in nanocrystalline multilayer thin film via grain boundary migration. The instability in such a film has been discussed on the basis of two interconnected phenomena : grain grooving and  PR instability. Grain grooving, which arises from the polycrystalline nature of the film that introduces a variation in curvature at  different regions of the film. This differential in curvature (perturbation) leads to the rupture of the film due to PR instability. The model results are also qualitatively consistent with our experimental observations. Though, beyond a critical thickness layers become stable, increasing the thickness of the multilayer thin film is always not a viable option to prevent film disintegration. As, increasing layer thickness also increase the grain size which may reduce the strength of the film by Hall-Petch effect \cite{0370-1301-64-9-303,petch1953cleavage}. We have shown through our phase-field model, reducing the bi-layer thickness using material with lower grain boundary mobility (i.e. higher melting point) will reduce the kinetics of film  disintegration. Making one of the layer amorphous or single crystal will be another way to prevent instability in the multilayer thin film. These findings are important for designing multialyer thin films for high temperature applications.

\section{Acknowledgment}
N.V. thanks Indian Institute of Science, Bangalore for the experimental facilities.

\bibliographystyle{elsarticle-num}
\bibliography{elsa}

\begin{thebibliography}{10}
\expandafter\ifx\csname url\endcsname\relax
  \def\url#1{\texttt{#1}}\fi
\expandafter\ifx\csname urlprefix\endcsname\relax\def\urlprefix{URL }\fi
\expandafter\ifx\csname href\endcsname\relax
  \def\href#1#2{#2} \def\path#1{#1}\fi

\bibitem{miyata2015high}
Y.~Miyata, K.~Nakayama, K.~Sugawara, T.~Sato, T.~Takahashi, High-temperature
  superconductivity in potassium-coated multilayer fese thin films, Nature
  materials 14~(8) (2015) 775--779.

\bibitem{hultman2000thermal}
L.~Hultman, Thermal stability of nitride thin films, Vacuum 57~(1) (2000)
  1--30.

\bibitem{xu2002microstructure}
J.~Xu, K.~Hattori, Y.~Seino, I.~Kojima, Microstructure and properties of crn/si
  3 n 4 nano-structured multilayer films, Thin Solid Films 414~(2) (2002)
  239--245.

\bibitem{yu2015multilayer}
S.~Yu, L.~Li, W.~Zhang, Z.~Sun, H.~Dong, Multilayer thin films with
  compositional pbzr0. 52ti0. 48o3/bi1. 5zn1. 0nb1. 5o7 layers for tunable
  applications, Scientific reports 5.

\bibitem{shih1992ti}
K.~Shih, D.~Dove, Ti/ti-n hf/hf-n and w/w-n multilayer films with high
  mechanical hardness, Applied physics letters 61~(6) (1992) 654--656.

\bibitem{verma2013detailed}
N.~Verma, V.~Jayaram, Detailed investigation of contact deformation in zrn/zr
  multiplayer—understanding the role of volume fraction, bilayer spacing, and
  morphology of interfaces, Journal of Materials Research 28~(22) (2013)
  3146--3156.

\bibitem{madan1996growth}
A.~Madan, X.~Chu, S.~Barnett, Growth and characterization of epitaxial mo/nbn
  superlattices, Applied physics letters 68~(16) (1996) 2198--2200.

\bibitem{madan1998enhanced}
A.~Madan, Y.-y. Wang, S.~Barnett, C.~Engstr{\"o}m, H.~Ljungcrantz, L.~Hultman,
  M.~Grimsditch, Enhanced mechanical hardness in epitaxial nonisostructural
  mo/nbn and w/nbn superlattices, Journal of applied physics 84~(2).

\bibitem{bozyazi2004comparison}
E.~Bozyaz{\i}, M.~{\"U}rgen, A.~F. {\c{C}}ak{\i}r, Comparison of reciprocating
  wear behaviour of electrolytic hard chrome and arc-pvd crn coatings, Wear
  256~(7) (2004) 832--839.

\bibitem{kapta2003degradation}
K.~Kapta, A.~Csik, L.~Daroczi, Z.~Papp, D.~Beke, G.~Langer, A.~Greer,
  Z.~Barber, M.~Kis-Varga, Degradation of ag/si multilayers during heat
  treatments, Vacuum 72~(1) (2003) 85--89.

\bibitem{moszner2016thermal}
F.~Moszner, C.~Cancellieri, M.~Chiodi, S.~Yoon, D.~Ariosa, J.~Janczak-Rusch,
  L.~Jeurgens, Thermal stability of cu/w nano-multilayers, Acta Materialia 107
  (2016) 345--353.

\bibitem{mukherjee2011thermal}
R.~Mukherjee, T.~Chakrabarti, E.~A. Anumol, T.~A. Abinandanan, N.~Ravishankar,
  Thermal stability of spherical nanoporous aggregates and formation of hollow
  structures by sintering a phase-field study, ACS nano 5~(4) (2011)
  2700--2706.

\bibitem{joshi2016phase}
C.~Joshi, T.~Abinandanan, A.~Choudhury, Phase field modelling of rayleigh
  instabilities in the solid-state, Acta Materialia 109 (2016) 286--291.

\bibitem{wang2016detachment}
F.~Wang, B.~Nestler, Detachment of nanowires driven by capillarity, Scripta
  Materialia 113 (2016) 167--170.

\bibitem{cahn1961spinodal}
J.~W. Cahn, On spinodal decomposition, Acta metallurgica 9~(9) (1961) 795--801.

\bibitem{allen1979microscopic}
S.~M. Allen, J.~W. Cahn, A microscopic theory for antiphase boundary motion and
  its application to antiphase domain coarsening, Acta Metallurgica 27~(6)
  (1979) 1085--1095.

\bibitem{fan1998phase}
D.~Fan, L.-Q. Chen, S.~Chen, P.~W. Voorhees, Phase field formulations for
  modeling the ostwald ripening in two-phase systems, Computational materials
  science 9~(3) (1998) 329--336.

\bibitem{poulsen2013three}
S.~O. Poulsen, P.~Voorhees, E.~M. Lauridsen, Three-dimensional simulations of
  microstructural evolution in polycrystalline dual-phase materials with
  constant volume fractions, Acta Materialia 61~(4) (2013) 1220--1228.

\bibitem{zhu1999coarsening}
J.~Zhu, L.-Q. Chen, J.~Shen, V.~Tikare, Coarsening kinetics from a
  variable-mobility cahn-hilliard equation: Application of a semi-implicit
  fourier spectral method, Physical Review E 60~(4) (1999) 3564.

\bibitem{frigo1998fftw}
M.~Frigo, S.~G. Johnson, Fftw: An adaptive software architecture for the fft,
  in: Acoustics, Speech and Signal Processing, 1998. Proceedings of the 1998
  IEEE International Conference on, Vol.~3, IEEE, 1998, pp. 1381--1384.

\bibitem{strutt1878instability}
J.~W. Strutt, L.~Rayleigh, On the instability of jets, Proc. London Math. Soc
  10~(4).

\bibitem{misra2005effects}
A.~Misra, R.~Hoagland, Effects of elevated temperature annealing on the
  structure and hardness of copper/niobium nanolayered films, Journal of
  materials research 20~(08) (2005) 2046--2054.

\bibitem{hoshen1976percolation}
J.~Hoshen, R.~Kopelman, Percolation and cluster distribution. i. cluster
  multiple labeling technique and critical concentration algorithm, Physical
  Review B 14~(8) (1976) 3438.

\bibitem{mullins1958effect}
W.~Mullins, The effect of thermal grooving on grain boundary motion, Acta
  metallurgica 6~(6) (1958) 414--427.

\bibitem{thornton1974influence}
J.~A. Thornton, Influence of apparatus geometry and deposition conditions on
  the structure and topography of thick sputtered coatings, Journal of Vacuum
  Science \& Technology 11~(4) (1974) 666--670.

\bibitem{singh2010residual}
D.~Singh, X.~Deng, N.~Chawla, J.~Bai, C.~Hubbard, G.~Tang, Y.-L. Shen, Residual
  stress characterization of al/sic nanoscale multilayers using x-ray
  synchrotron radiation, Thin Solid Films 519~(2) (2010) 759--765.

\bibitem{karim2007influence}
S.~Karim, M.~Toimil-Molares, W.~Ensinger, A.~Balogh, T.~Cornelius, E.~Khan,
  R.~Neumann, Influence of crystallinity on the rayleigh instability of gold
  nanowiresthis paper is dedicated to professor dr h fue{\ss} on the occasion
  of his 65th birthday., Journal of Physics D: Applied Physics 40~(12) (2007)
  3767.

\bibitem{0370-1301-64-9-303}
E.~O. Hall, \href{http://stacks.iop.org/0370-1301/64/i=9/a=303}{The deformation
  and ageing of mild steel: Iii discussion of results}, Proceedings of the
  Physical Society. Section B 64~(9) (1951) 747.
\newline\urlprefix\url{http://stacks.iop.org/0370-1301/64/i=9/a=303}

\bibitem{petch1953cleavage}
N.~Petch, The cleavage strength of polycrystals, J. Iron Steel Inst. 174 (1953)
  25--28.

\end{thebibliography}

\end{document}